\documentclass{article}
\usepackage[margin=1in]{geometry}	
\usepackage{amsfonts,amsmath,amssymb,wasysym}	
\usepackage{fancyhdr} 							
\usepackage[explicit]{titlesec}	
\usepackage{indentfirst}						
\usepackage[round]{natbib}						
\usepackage{graphicx}
\usepackage{caption}
\usepackage{hyperref}
\usepackage{setspace}
\usepackage[titles]{tocloft}
\usepackage{pdfpages} 
\usepackage{bibentry}
\usepackage{algorithm2e}
\usepackage{amsthm, mathtools}
\usepackage[most]{tcolorbox}
\usepackage[symbol]{footmisc}

\newcommand{\V}[1]{\mathbf{#1}} 
\newcommand{\B}{\mathbf{B}}

\newcommand{\gkeyll}{{\tt Gkeyll}}

\DeclarePairedDelimiter{\norm}{\lVert}{\rVert}

\title{Multi-Point Gradient Estimation in Turbulence}
\author{Theodore Broeren, Kristopher Klein}
\date{\today}

\begin{document}
\maketitle

\begin{abstract}
When studying turbulence, it is often desirable to be able to estimate the local spatial gradient of a vector quantity using in situ measurements from a small number of irregularly spaced points. While previous studies have focused on the accuracy of these methods as the number of measurement points varies, we focus on the accuracy of gradient estimations as a function of shape of the measurement point configuration. We find that for well-shaped configurations that are of correct relative size, we can very accurately estimate the local spatial gradient using only four points. 
\end{abstract}

\section{Introduction}

Previous work has analyzed the accuracy of spatial gradient estimation techniques in the domain of fluid turbulence \citep{watanabe2017}. This analysis utilized a method of estimating the local spatial gradient via a small number of sample points in three dimensions, called least-squares gradient computation \citep{DeKeyser:2007}. This method was developed to estimate gradients in space plasmas utilizing in-situ measurements from multi-spacecraft observatories. However, this estimation method can also be applied to numerical turbulence simulations where sample points are irregularly spaced, such as simulations written within a Lagrangian framework.

\section{Methods}
The previous study \citep{watanabe2017} found that for a randomly generated set of sample points distributed in three-dimensions, the number of points included in the least-squares gradient computation greatly effects the accuracy of the local gradient estimation. This was determined by comparing the coarse grained velocity tensor (serving as the baseline true local gradient) to the spatial gradient computed via least-squares regression.

\subsection{Least-Squares Gradient Computation}
A multi-point method of estimating the spatial gradient of a vector field is least-squares gradient computation. This method formulates an optimization problem \citep{Harvey:1998}, where we wish to minimize the expression 
\begin{equation}
	\sum_{n=1}^3 \sum_{\alpha=1}^N \sum_{\beta=1}^N \left[\frac{\partial B_{n}}{\partial r_m} \left( r_{m}^{(\alpha)} - r_{m}^{(\beta)} \right) -  \left( B_{n}^{(\alpha)} - B_{n}^{(\beta)} \right) \right]^2. \label{eqn:least_sq_min}
\end{equation}
The solution of this minimization problem is found to be
\begin{equation}
	\frac{\partial B_{n}}{\partial r_l} = \frac{1}{N^2} \left[ \sum_{\alpha \neq \beta} \left( B_{n}^{(\alpha)} - B_{n}^{(\beta)} \right) \left( r_{k}^{(\alpha)} - r_{k}^{(\beta)} \right) \right] \overline{\overline{R}}_{kl}^{-1},
\end{equation}
using the critical points of Eqn \ref{eqn:least_sq_min}. The sum $\sum_{\alpha \neq \beta}$ in this equation is over all $N(N-1)/2$ independent terms where $\alpha \neq \beta$ and $\overline{\overline{R}}$ refers to the volumetric tensor (defined in Eqn \ref{eqn:R_tensor}).

Data from configurations containing $N \geq 4$ sample points can be incorporated using this formulation, however this least-squares formulation will always result in a single 'linear' estimate of the local gradient. Some spacecraft applications have combined this estimation with self-consistency checks to provide a method of automatic error estimation, as well as Taylor's hypothesis to increase the number of spatial samples \citep{DeKeyser:2008}. By using a distance threshold, one can estimate a local linear gradient that varies as a function of space at scales smaller than the separations of a static spacecraft configuration \citep{Hamrin:2008}.

\subsection{Radial Basis Function Gradient Computation}
A radial basis function (RBF), $\varphi(\norm{\mathbf{r}^{(n)} - \mathbf{r}})$, is a function whose value depends only on the distance between the input, $\mathbf{r}$, and some fixed point, $\mathbf{r}^{(n)}$, called a center. We use a weighted sum of these RBFs to interpolate $\B$ at any arbitrary position $\mathbf{r}$ using measurements made at the $N$ sample point locations:
\begin{equation}
	B_m(\mathbf{r}) = \sum_{i=1}^{N} \sum_{j=1}^{T} a_{n} \varphi(\norm{\mathbf{r}^{(n)} - \mathbf{r}}) \hspace{0.5cm} \forall m \in \{x,y,z\}.\label{eqn:RBF}
\end{equation}
To perform this interpolation, we must pick a functional form $\varphi$ and we must learn the values of the coefficients $a_{n}$. The estimated values of the coefficients are learned through weighted linear regression, using the measured values of vector field and sample point position as a dataset to be fit to. This method is easily implemented in Python using the Scipy interpolate packages \textit{Rbf} or \textit{RBFInterpolator} \citep{SciPy:2020, Fasshauer:2007}.

Several smooth RBF kernel functions are commonly used in the literature including Gaussian, multiquadric, and inverse quadric \citep{Fasshauer:2007}. We tested these three options and found that the multiquadric kernel performed the best for multi-spacecraft configurations measuring magnetic fields of plasma turbulence in situ \citep{Broeren:2024}. The multiquadric RBF kernel has the form
\begin{equation}
	\varphi(\norm{\mathbf{r}^{(i,j)}-\mathbf{r}}) = \sqrt{1 + \left(\norm{\mathbf{r}^{(i,j)}-\mathbf{r}}/\sigma_{\text{rbf}} \right)^2}. \label{eqn:MQ_RBF}
\end{equation}
In this equation, $\sigma_{\text{rbf}}$ is a tunable hyper-parameter that is proportional to the radius of influence that measured data points have on the interpolation. For this interpolation we set this parameter as a function of sample point configuration size $\sigma_{\text{rbf}} = L/2$. Once the vector field is interpolated, we then compute the course grained velocity tensor of the gradient of this interpolated field, which is computed using a finite difference method.

\begin{figure}[t]
	\centering
	\textbf{Course Grained Velocity Tensor}\par\medskip
 \includegraphics[trim={1.2in 0.6in 0.6in 0.75in}, clip, width=0.60\textwidth]{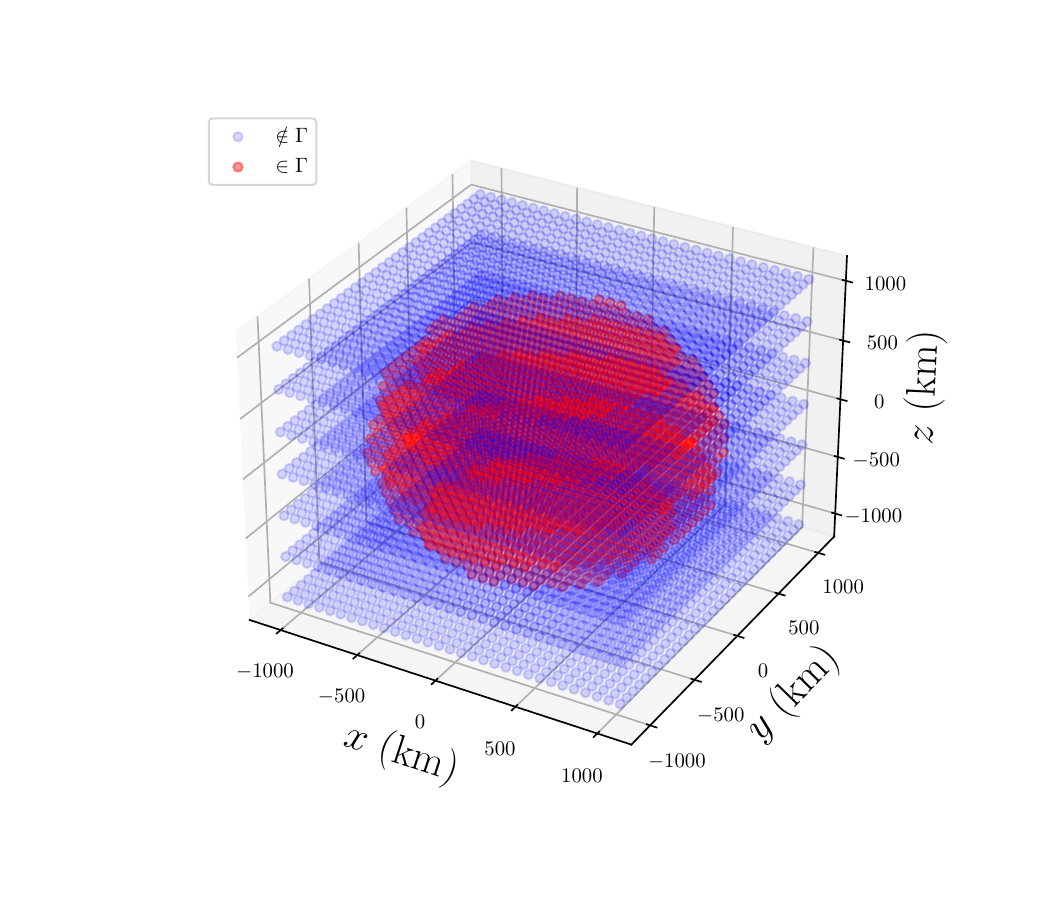}
        \caption{An example set of points that is used to compute the course grained velocity tensor. The gradient is computed using a finite difference method on all blue points. Then, the course grained velocity tensor is computed by averaging over all red points, within the sphere $\Gamma$ (diameter $L=2000$ km).}
	\label{fig:Velocity_Tensor} 
\end{figure}

\subsection{Gradient Accuracy}
The course grained velocity tensor is defined as
\begin{equation}
    \partial_i B_m = \int_{\Gamma} \partial_i B_m dx \label{eqn:vel_tensor}
\end{equation}
where $\Gamma$ is a sphere of diameter $L$ (defined in Eqn \ref{eqn:L}). As we are using a numerical simulation for our baseline comparison, the derivative terms on the right-hand side of equation \ref{eqn:vel_tensor} are estimated using a second-order finite difference scheme. We then find the course grained velocity tensor of scale $L$ by averaging the derivatives over the region within $\Gamma$. See Figure \ref{fig:Velocity_Tensor} for pictorial representation of $\Gamma$.

We compute the course grained velocity tensor of the simulations vector field. We then compare this to the course grained velocity tensor computed from a vector field reconstructed using either the least-squares or RBF methods. This comparison is done using the Pearson product-moment correlation coefficient. The correlation coefficient between two variables, $x$ and $y$, can be computed for a finite sample of $n$ pairs of data points as
\begin{equation}
    R = \frac{ \sum_{i=1}^n\left( x_i - \bar{x} \right) \sum_{i=1}^n\left( y_i - \bar{y} \right) }{ \sqrt{\sum_{i=1}^n\left( x_i - \bar{x} \right)^2} \sqrt{\sum_{i=1}^n\left( y_i - \bar{y} \right)^2} } \in \left[ -1, 1\right],
\end{equation}
where $\bar{x} = \sum_{i=1}^n x_i $ is the sample mean. Intuitively, this coefficient will be near its maximal value of $+1$ when the true value of the gradient ($x$) and the estimated value of the gradient ($y$) exhibit similar behavior over a number of realizations ($n$). This coefficient will be near 0 if the gradient computed from the simulation does not correlate to the gradient extracted from the reconstructed vector field.

\subsection{Turbulence Simulation}
The turbulence simulation chosen is designed to represent plasma behavior in the pristine solar wind at 1AU. We utilize the open source plasma physics code \gkeyll\footnote{\gkeyll\ (\url{https://gkeyll.readthedocs.io/en/latest/}) is the product of a multi-institution collaboration, led by Princeton University and the Princeton Plasma Physics Laboratory. It contains solvers for gyrokinetic equations, Vlasov-Maxwell equations, and multi-fluid equations. } to run this simulation. The input file for the \gkeyll\ simulation used in this work is available in the Zenodo repository \citep{broeren:2024_zenodo}. In particular, we utilize the magnetic fields extracted from a proton-electron fluid simulation evolved using a five (1 density, 3 velocity,  1 energy) moment multi-fluid model \citep{Wang:2015}.

Our simulation uses a reduced proton to electron mass ratio of $m_i / m_e = 100$, a temperature ratio of $T_i / T_e = 1$, an Alfvén velocity of $v_{A} = 0.02 c$, and a plasma beta of $\beta_i = 1$. We use the adiabatic equation of state with an adiabatic index $\gamma = 5/3$. As the background magnetic field is chosen to be uniform along the $\hat{z}$ direction, $\V{B}_0 = B_0 \hat{\V{z}}$, the simulation domain is elongated 
\begin{equation}
	L_x = L_y = 0.2 L_z = 100 \pi \rho_i
\end{equation}
in this direction to compensate for the preferential cascade of energy along directions perpendicular to it. The simulation domain has a resolution of $n_x \times n_y \times n_z = 448 \times 448 \times 448$ and lengths are normalized to the proton gyroradius $\rho_i = 100$km. For more details on this simulation, see our previous works that utilize the same simulation \citep{Broeren:2021, Broeren:2024, broeren2024_phd}.

\subsection{Shape of Configuration}
We wish to systematically describe the 'well-shapedness' of a sample point configuration in 3D space using a single scalar quantity. We note that sample points that are arranged on the same plane or line will not be able to measure gradients along one or two spatial directions. To quantify how sensitive configurations of sample points are to fluctuations along all directions, the dimensionless quantities elongation and planarity, formally defined below, are commonly used in multi-spacecraft analysis. These quantities were originally derived to describe the geometry of four-spacecraft configurations. We can simplify the commonly used quantities of elongation and planarity into a single scalar shape parameter to describe how well-suited a configuration of sample points is to fluctuations along all directions. This single scalar quantity will be a convenient way to describe the geometry of a configuration of sample points by compressing the information contained in the set of all $N$ ($\hat{x}$, $\hat{y}$, $\hat{z}$) values.

Following the work laid out in Chapter 12 of \cite{Paschmann:1998}, we start our derivation of elongation and planarity by defining the barycenter, $\mathbf{r}_0$, of the $N$ spatial points as
\begin{equation}
	\mathbf{r}_0 = \frac{1}{N} \sum_{n=1}^N \mathbf{r}^{(n)} ,
\end{equation}
where $\mathbf{r}^{(n)}$ is the position vector for point $n$. We then use the barycenter to define the $3\times3$ volumetric tensor matrix
\begin{equation}
	\overline{\overline{R}}_{jk} = \frac{1}{N} \sum_{n=1}^N \left[  r_j^{(n)} - (r_0)_{j} \right]  \left[  r_k^{(n)} - (r_0)_{k} \right] , \label{eqn:R_tensor}
\end{equation}
where $ r_j^{(n)}$ is the $j^{th}$ component of the $n^{th}$ point's position. 

We take the square-roots of the volumetric tensor's eigenvalues to find its singular values $a \geq b \geq c$. These singular values describe the 3 semi-axis of an ellipsoid that approximates the size and orientation of the sample point configuration. We use these singular values to define elongation ($E$) and planarity ($P$), as well as the characteristic size ($L$) of the configuration.


\textbf{Elongation} is computed using the ratio of the longest two semi-axes of the ellipsoid
\begin{equation}
	E = 1 - b/a \in [0,1]. \label{eqn:E}
\end{equation}
It approaches its maximal value for configurations that are nearly co-linear.

\textbf{Planarity} compares the smallest two semi-axes of the ellipsoid
\begin{equation}
	P = 1 - c/b \in [0,1]. \label{eqn:P}
\end{equation}
It approaches its maximal value for configurations that are nearly co-planar.

\textbf{Characteristic size} describes the overall size of the ellipsoid along its semi-major axis
\begin{align}
	L &= 2a \in [0,\infty). \label{eqn:L}
\end{align}

\begin{figure}[t!]
	\centering
	\textbf{Least-Squares Gradient Estimation}\par\medskip
	\begin{tabular}{cc}
		\includegraphics[trim={0.2in 0.0in 0.0in 0.10in}, clip, width=0.45\textwidth]{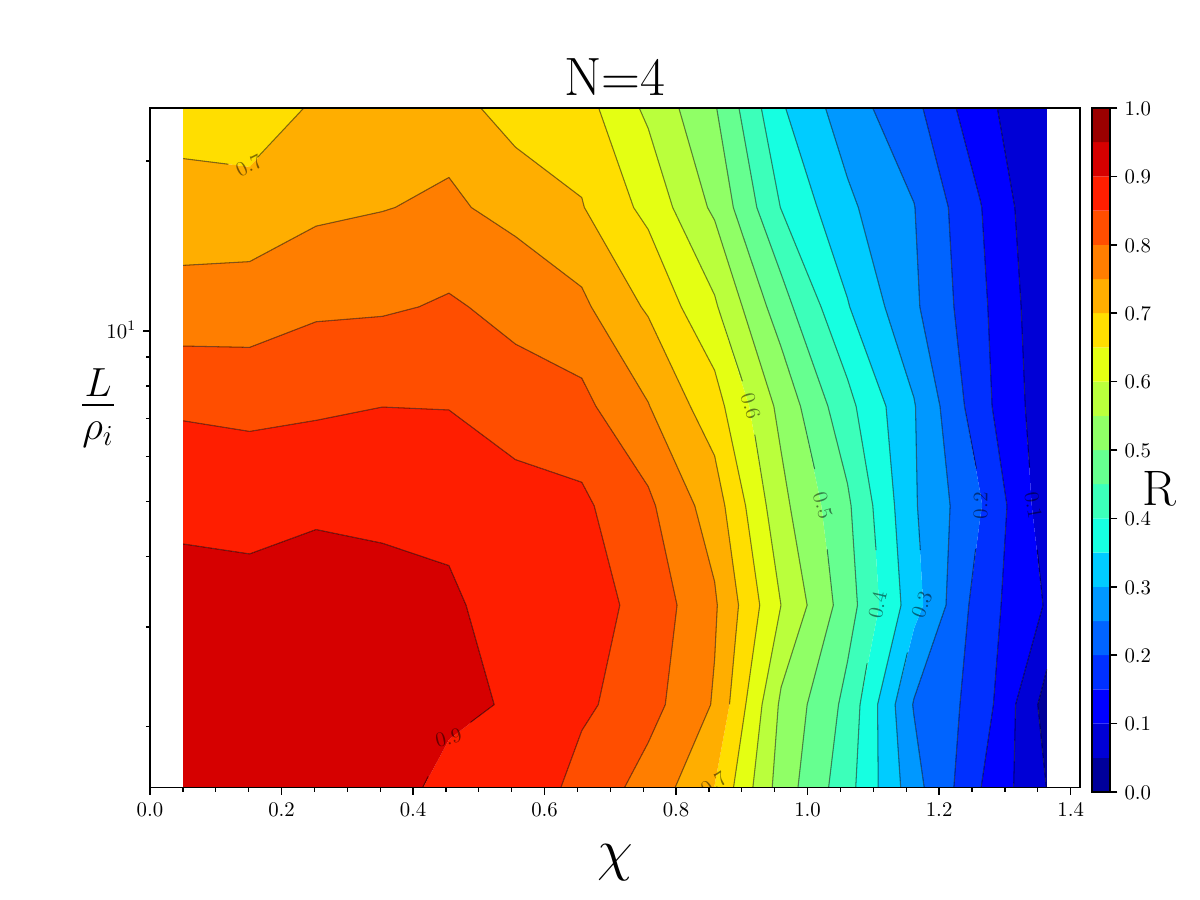} &
		\includegraphics[trim={0.2in 0.0in 0.0in 0.10in}, clip, width=0.45\textwidth]{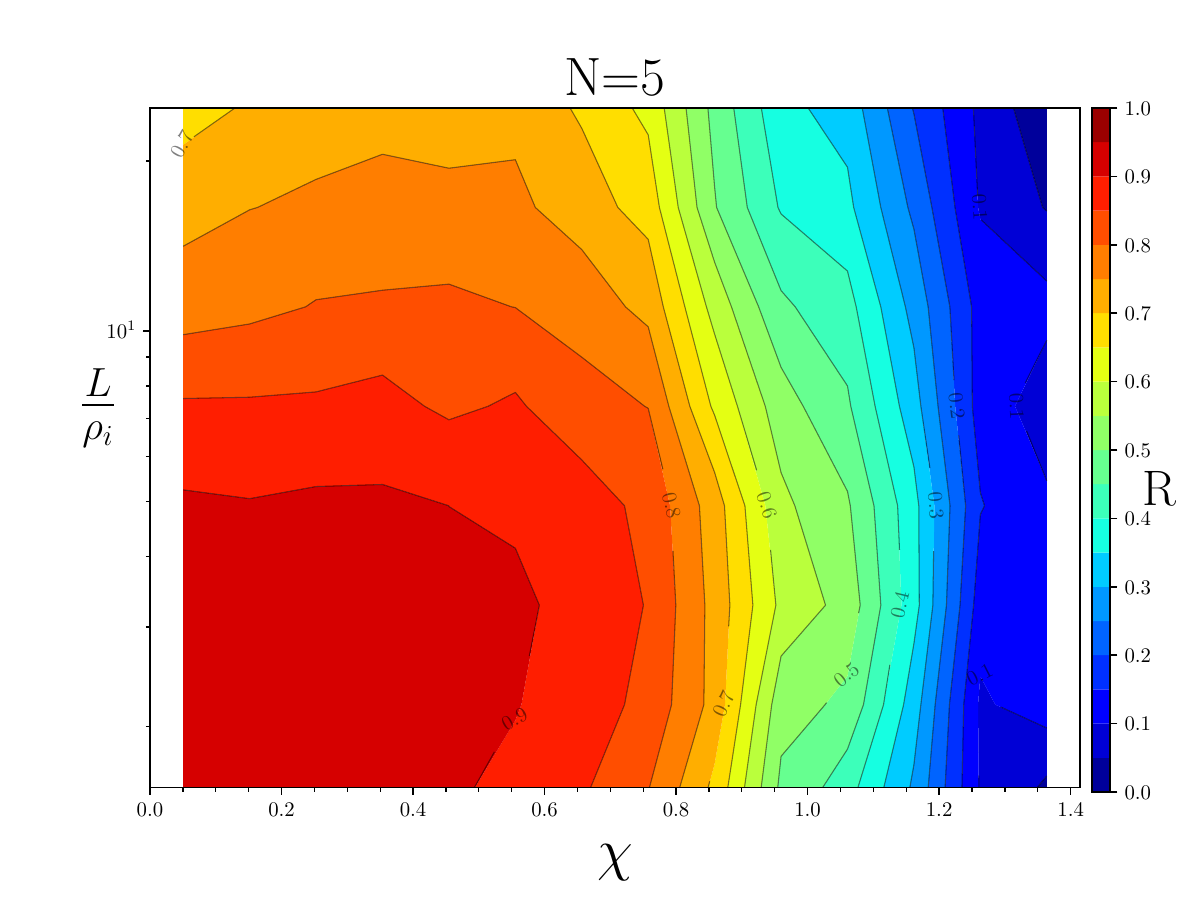} \\
		\includegraphics[trim={0.2in 0.0in 0.0in 0.10in}, clip, width=0.45\textwidth]{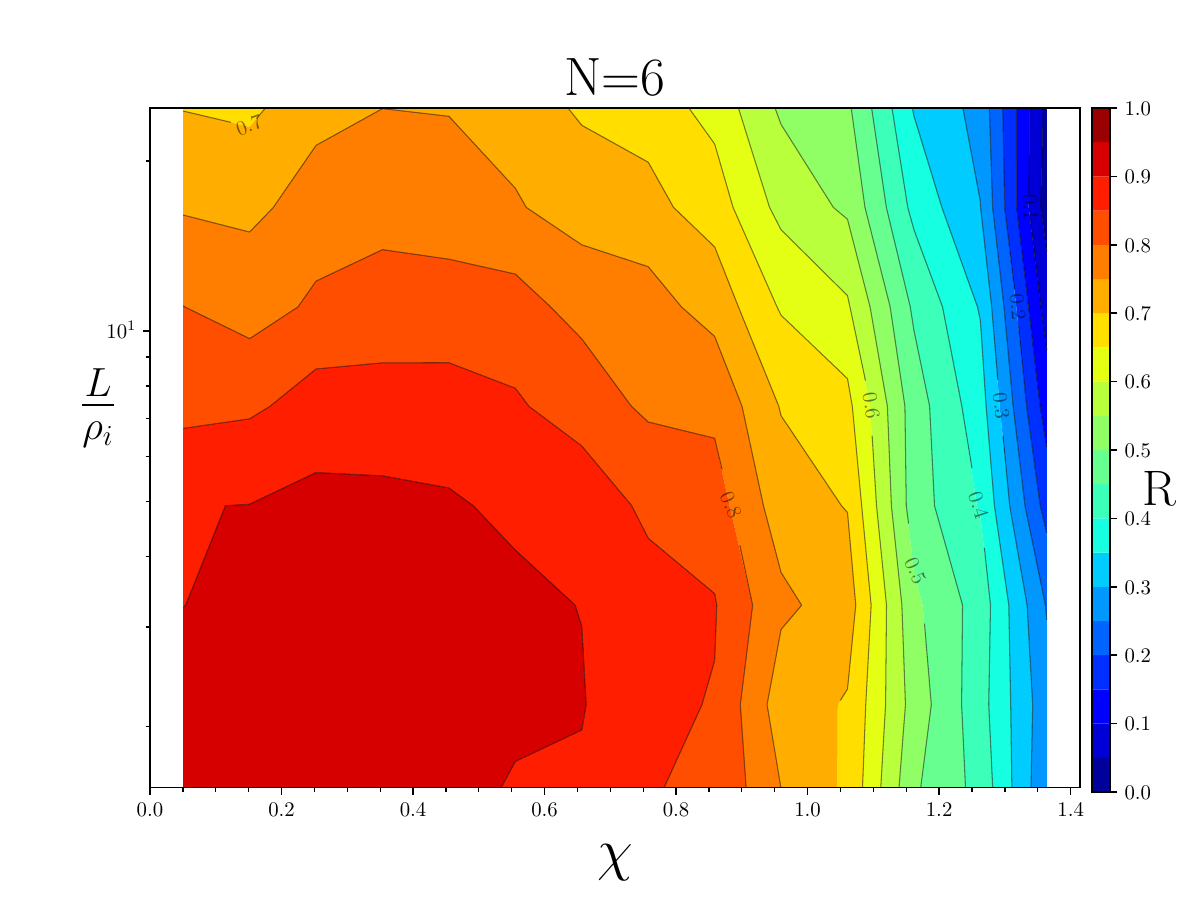} &
		\includegraphics[trim={0.2in 0.0in 0.0in 0.10in}, clip, width=0.45\textwidth]{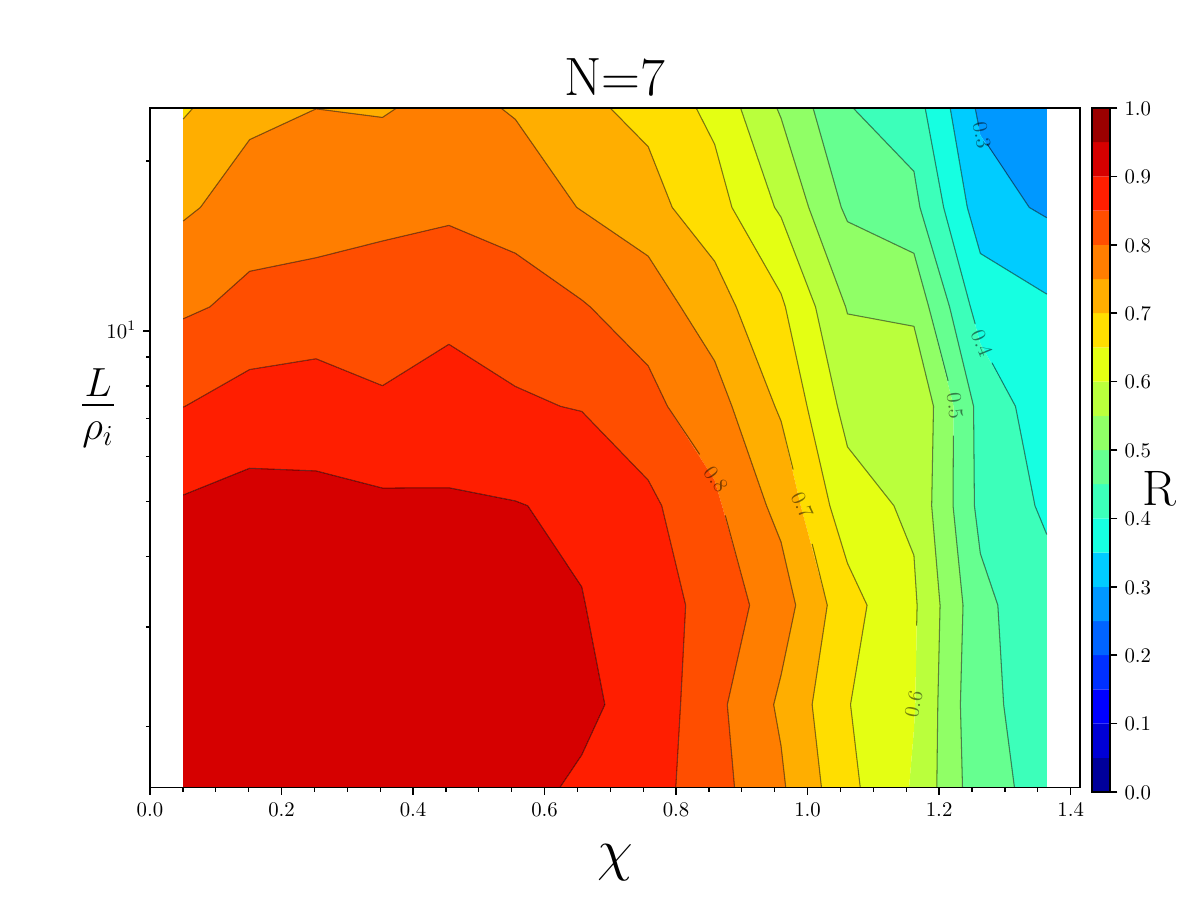} \\
		\includegraphics[trim={0.2in 0.0in 0.0in 0.10in}, clip, width=0.45\textwidth]{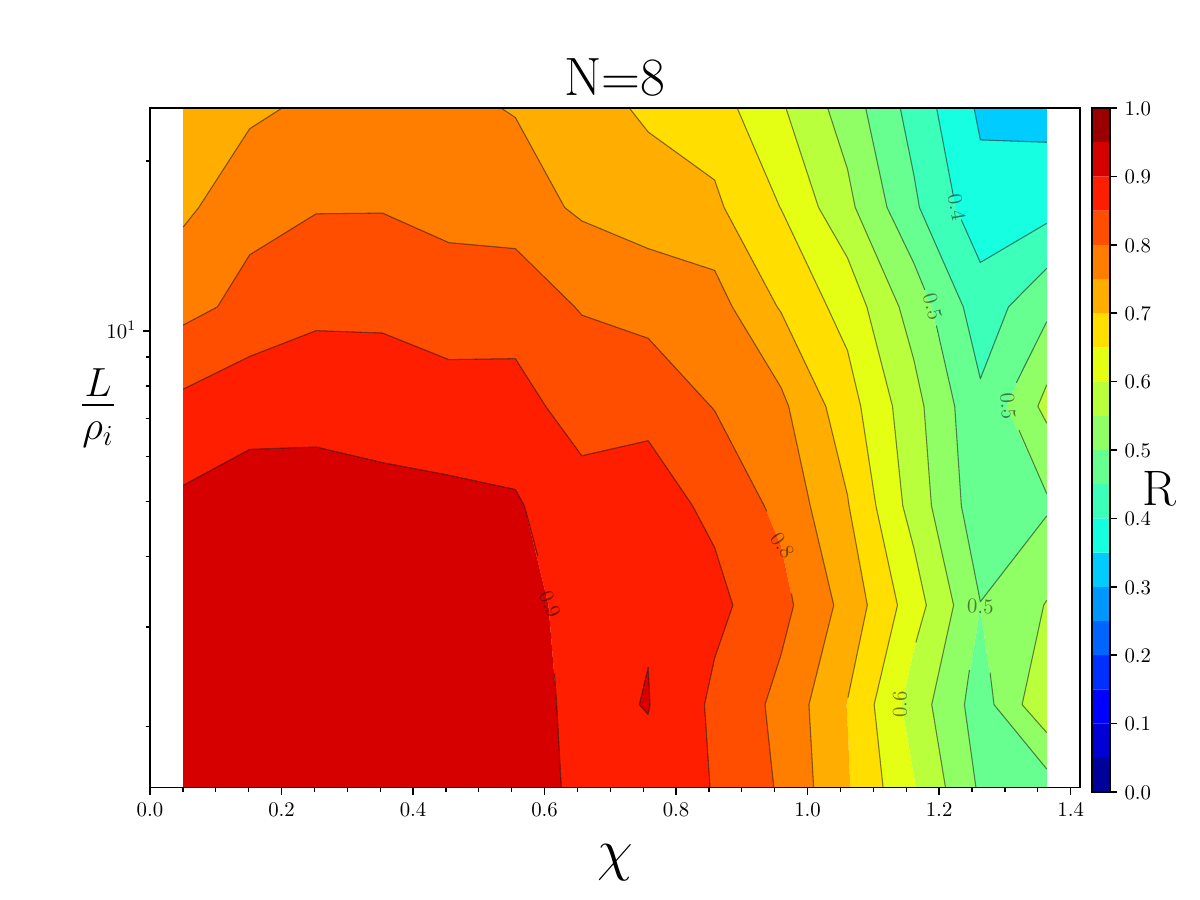} &
		\includegraphics[trim={0.2in 0.0in 0.0in 0.10in}, clip, width=0.45\textwidth]{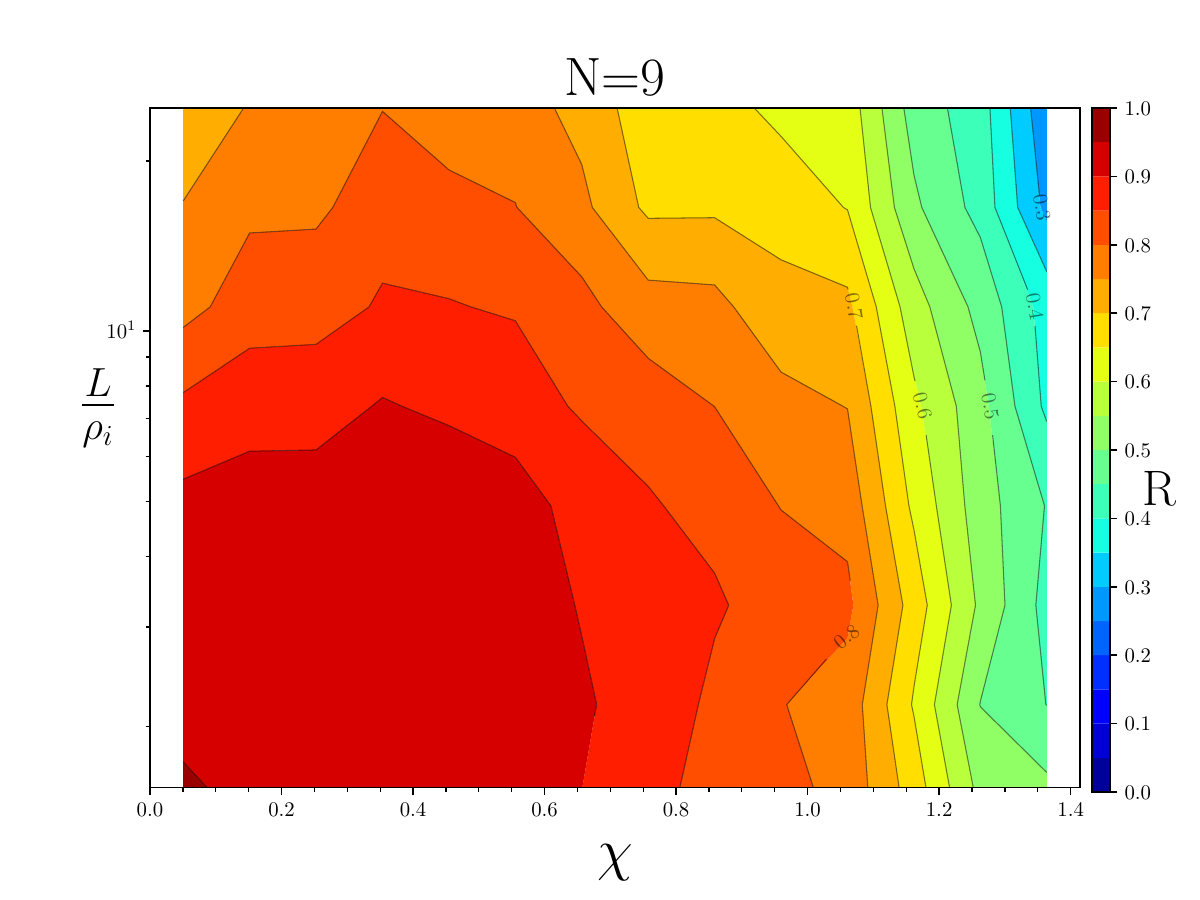}
	\end{tabular}
	\caption{We use many configuration in a simulation of turbulence to quantify the average correlation (R) between the true spatial gradient (estimated using a second-order finite difference on the simulation grid) and the gradient estimated using the least-squares estimation technique. The measurement point configurations have characteristic sizes $L \in [2 \Delta x, 4000]$km where $\Delta x = 70.1$km. The data is smoothed over a step-size of $\Delta \chi = 0.1$.}
	\label{fig:LS_grad_corr} 
\end{figure}

Previous studies of these quantities have shown that the well-shapedness of a configuration is related to elongation and planarity symmetrically (see page 343 of \cite{Paschmann:1998} or \cite{Broeren:2021}). Therefore, we define a single shape parameter that combines elongation and planarity
\begin{equation}
	\chi = \sqrt{E^2 + P^2} \in [0,\sqrt{2}]. \label{eqn:chi}
\end{equation}
This new parameter, $\chi$, is near its minimal value for well-shaped 'spherical' configurations of sample points and near its maximal value for poorly-shaped (co-linear and/or co-planar) configurations.

\section{Results}
To study the effect of sample point configuration shape on the accuracy of the estimated gradient, we use the dataset of randomly generated sample point configurations that was described in \cite{Broeren:2023}. This dataset contains roughly 300 configurations of each number of sample points $N \in \{4,5,6,7,8,9\}$. The configurations are generated to uniformly span the possible shape parameter regime $\chi \in (0,\sqrt{2})$, while simultaneously spanning possible of values elongation and planarity $E,P \in (0,1)$.

\subsection{Least-Squared Gradient}
We place each of the over 300 sample point configurations into the numerical simulation of turbulence at 20 locations. At each of these locations, we also scale each sample point configuration so that it attains 30 different characteristic sizes $L$. For each of these sample point placement and configuration combinations, we use the least-squares gradient computation method to estimate the local gradient of the vector magnetic field utilizing synthetic in situ measurements. We compare this estimated gradient to the true gradient found from the grid of vector magnetic field values drawn from the turbulence simulation itself. By repeating this Monte-Carlo approach for sample configurations containing $N \in \{4,5,6,7,8,9\}$ points, we generate the results presented in Figure \ref{fig:LS_grad_corr}.

For all numbers of points we see that for small scales ($L/\rho_i$ approaching 1), the estimated gradient closely matches the gradient drawn from the simulation itself. As the scale size increases, we see that the gradient estimation degrades following an approximately a log-linear trend $R \propto \log(L/\rho_i)^{-1}$.

We also see that if relatively few sample points are used in the gradient estimation $4 \leq N \leq 6$, there is a steep performance drop-off at some threshold $\chi(N)$ value. From these experiments, it appears that if 7 or more points are utilized, then this performance degradation can be avoided for all but the most misshapen configurations.

\subsection{Radial Basis Function Gradient}
\begin{figure}[t!]
	\centering
	\textbf{Radial Basis Function Gradient Estimation}\par\medskip
	\begin{tabular}{cc}
		\includegraphics[trim={0.2in 0.0in 0.0in 0.10in}, clip, width=0.45\textwidth]{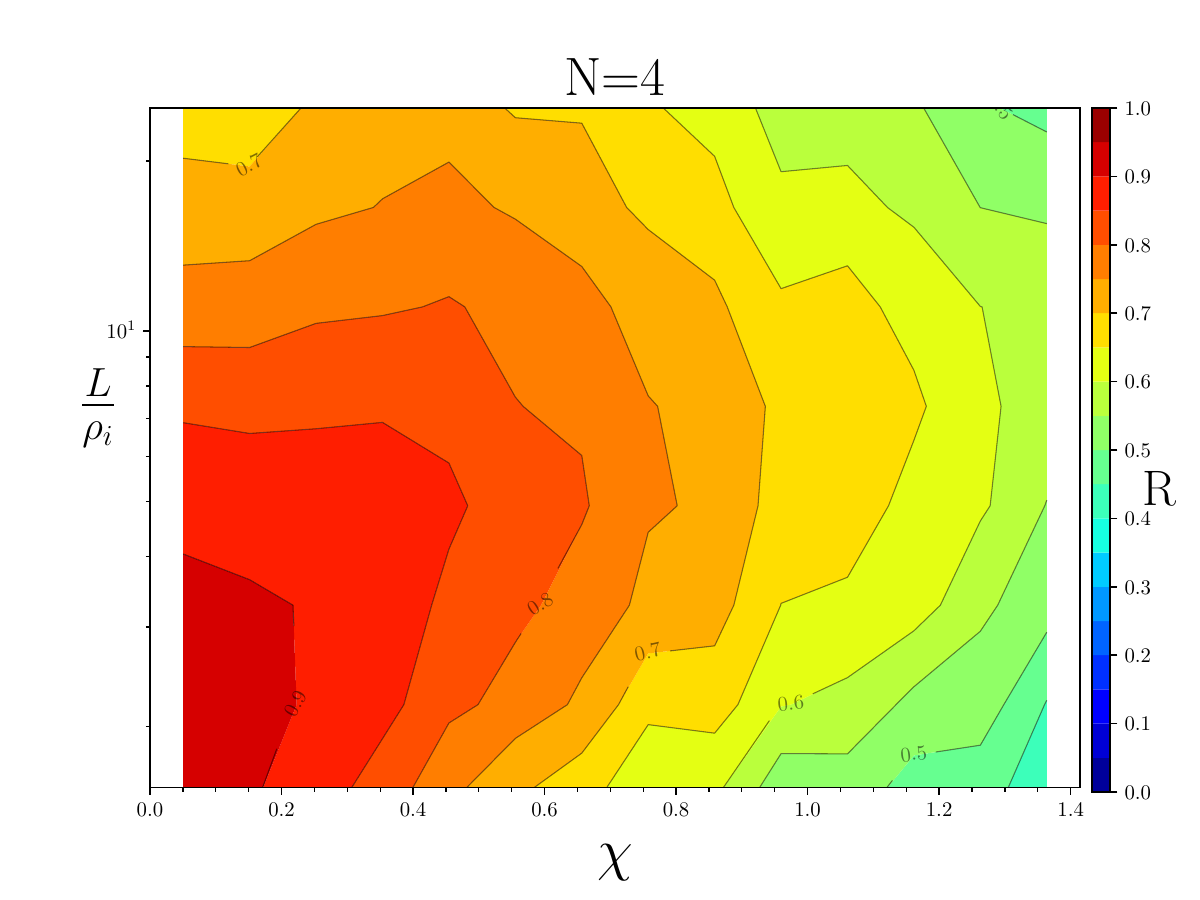} &
		\includegraphics[trim={0.2in 0.0in 0.0in 0.10in}, clip, width=0.45\textwidth]{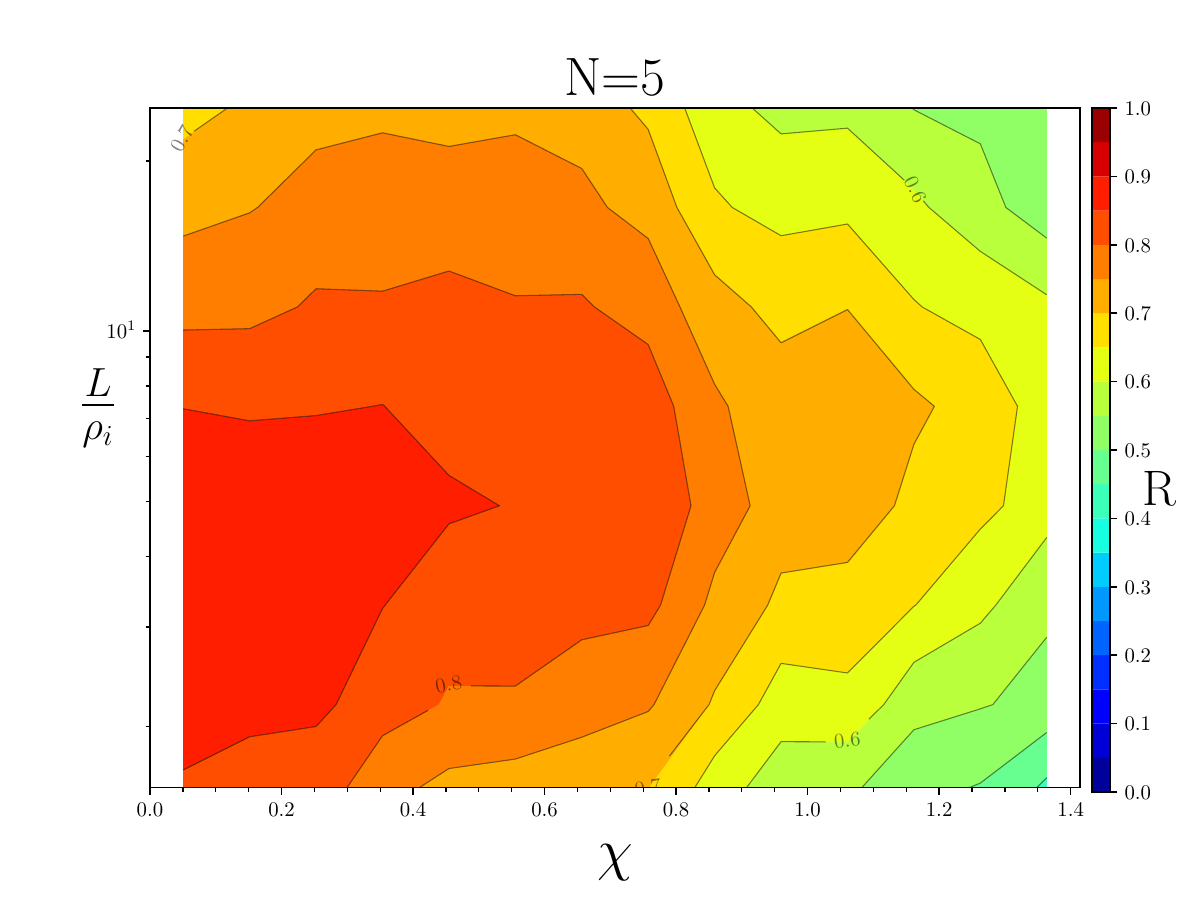} \\
		\includegraphics[trim={0.2in 0.0in 0.0in 0.10in}, clip, width=0.45\textwidth]{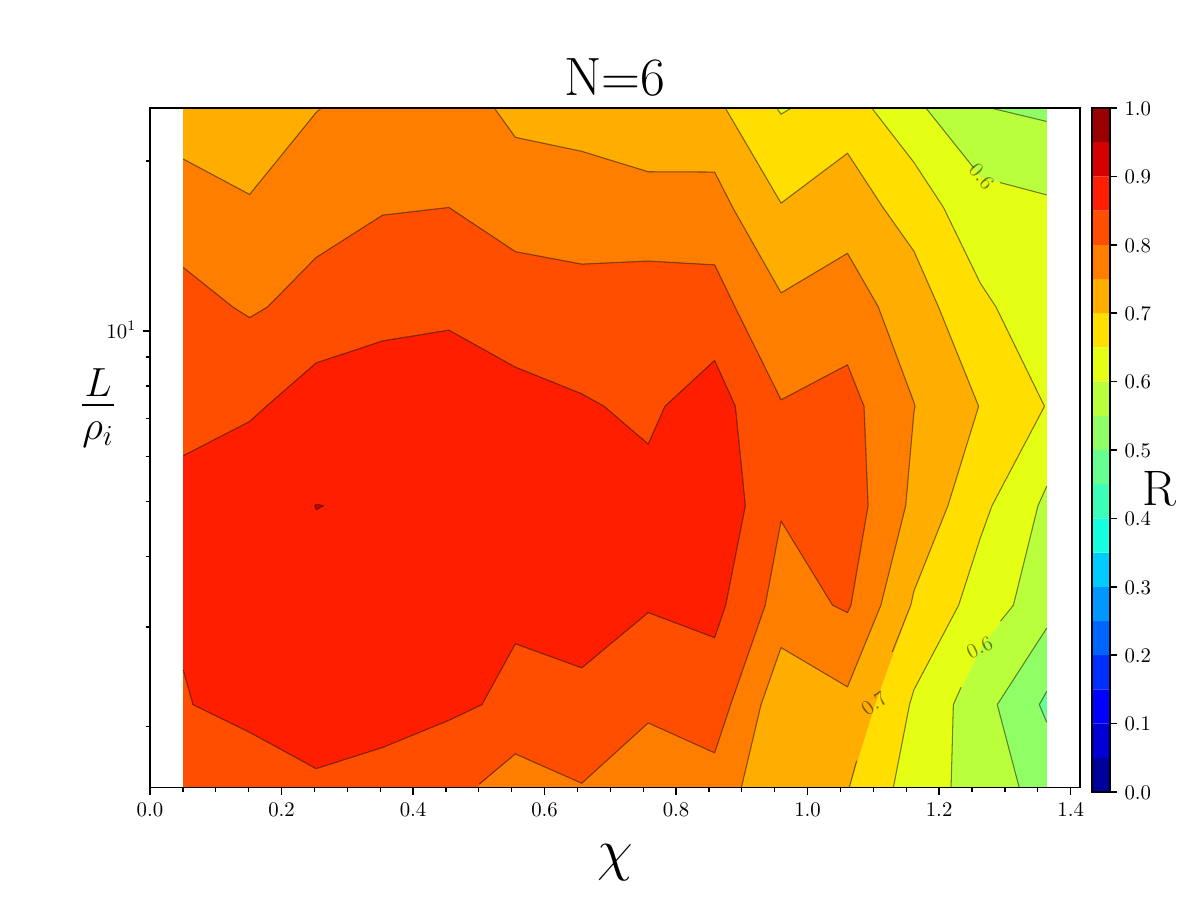} &
		\includegraphics[trim={0.2in 0.0in 0.0in 0.10in}, clip, width=0.45\textwidth]{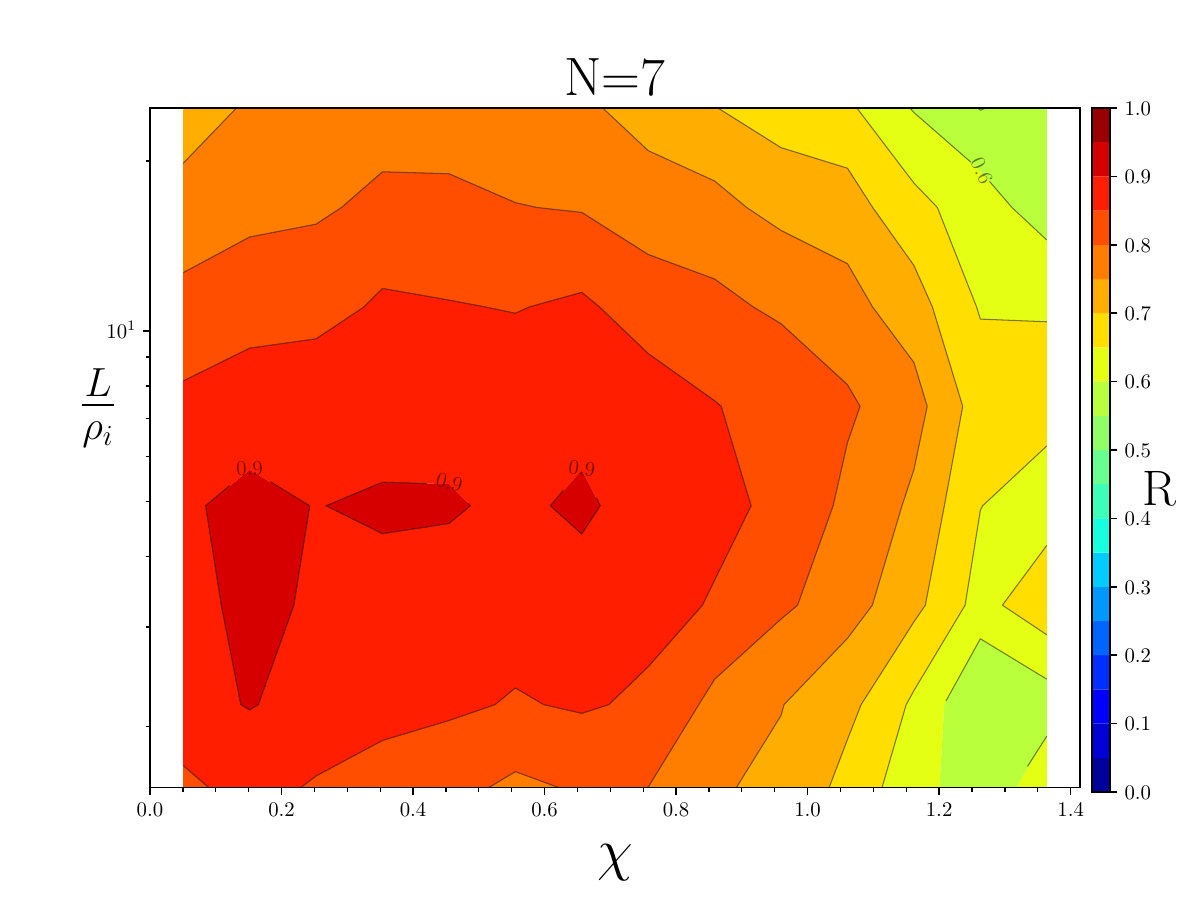} \\
		\includegraphics[trim={0.2in 0.0in 0.0in 0.10in}, clip, width=0.45\textwidth]{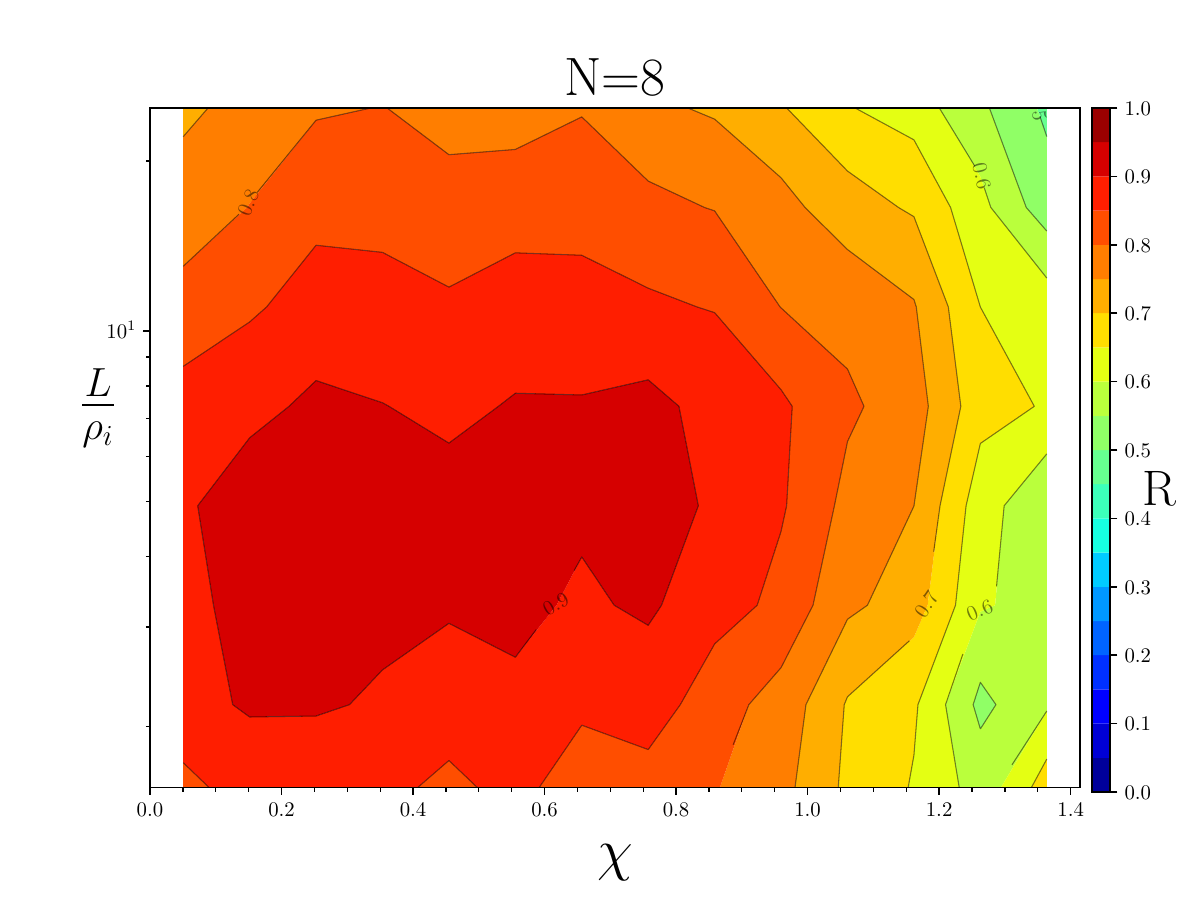} &
		\includegraphics[trim={0.2in 0.0in 0.0in 0.10in}, clip, width=0.45\textwidth]{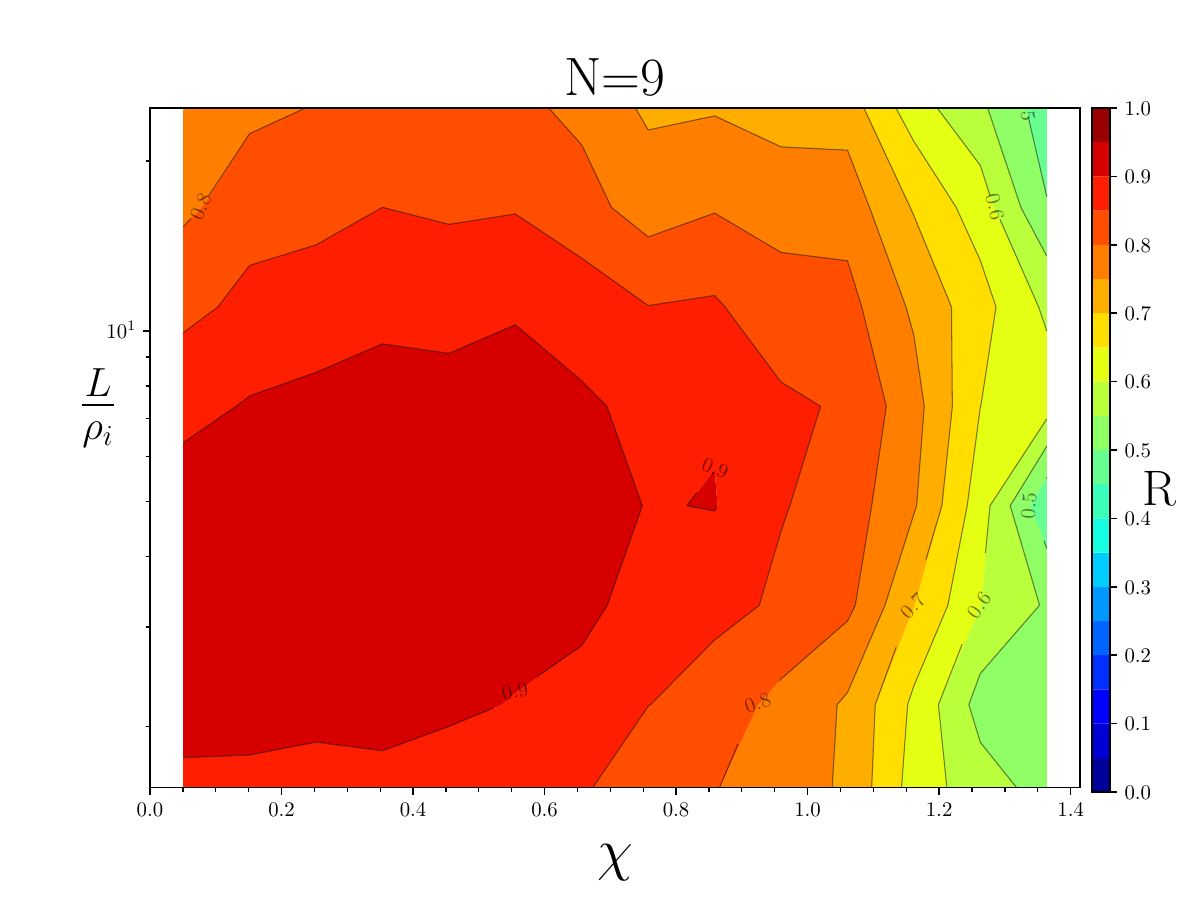}
	\end{tabular}
	\caption{We quantify the average correlation (R) between the true spatial gradient and the gradient estimated using the radial basis function estimation technique. The panels data is displayed identically to Figure \ref{fig:LS_grad_corr}.}
	\label{fig:rbf_grad_corr} 
\end{figure}

We perform the same Monte-Carlo experiment for the radial basis function method that was done for the least-squares gradient. Results obtained using the radial basis function method are displayed in Figure \ref{fig:rbf_grad_corr}. From examining the result of the $N=4$ point experiment, we see that the gradient estimation gathered from the radial basis function method is not as effected by high $\chi$ values. There appears to be no threshold value of $\chi$ that separates well and poorly performing configuration. Instead, there is a more gradual decrease in performance as $\chi$ increases. As the number of points in each configuration $N$ increases, the range of configuration shapes that are very good at estimating the gradient widens significantly.

We see a similar scaling of gradient accuracy with scale size that was present when applying the lease-squares gradient method. However, we do see a shift in scale space where the region of highest accuracy is located. This may tell us that the characteristic size of a configuration of points using the RBF method should be thought of in a slightly different manner than it is when using the least-squares method.

\section{Discussion}
We have seen that a non-uniform configuration of four points can be used to estimate the spatial gradient of a turbulent vector field. The accuracy of this estimation depends on both the scale size of the fluctuation compared to the configuration of points and the geometrical shape ($\chi$) of the configuration. Gradient estimations made using different methods will depend on the shape in different ways. However, we have observed that the relative size of the configuration will be key, independent of the gradient estimation technique that is employed. 

These findings have consequences in both computational and applied physics. As turbulence is a multi-scale phenomena, one can design 'well-shaped' configurations of points that are at slightly differing scales to study what the spatial gradients will be at all scales simultaneously. Knowledge of how accurate one's estimation of the spatial gradient is will be essential in the design of these different 'well-shaped' multi-scale configurations. This will allow physicists to deploy an overall fewer number of points that are equally as effective at studying turbulent phenomena. This fewer number of points could be the difference between a discrete particle simulation being executable or not, or a multi-spacecraft mission (such as \cite{klein:2023}) being over or under budget.

\section*{Acknowledgments}
This research was supported by the International Space Science Institute in Bern, through ISSI International Team project \#556 \href{https://teams.issibern.ch/energtransferspaceplasmas/}{Cross-Scale Energy Transfer in Space Plasmas}.

\bibliographystyle{abbrvnat} 
\bibliography{bibliography_urls.bib}

\end{document}